\documentclass[prl,nopacs,twocolumn,preprintnumbers,notitlepage,amsmath,amssymb,superscriptaddress]{revtex4-1}
\usepackage{amssymb}
\usepackage{amsmath}

\usepackage{color}
\usepackage[pdftex]{graphicx}

\usepackage{mathtools}

\usepackage{makecell}

\usepackage{enumerate}
\usepackage{mathtools}
\usepackage{stmaryrd}

\usepackage{enumitem}

\usepackage{textcomp}
\usepackage{gensymb}

\newcommand{\ex}[1]{\mathrm{e}^{#1}}

\newcommand{\dd}[0]{\mathrm{d}}
\newcommand{\ii}[0]{\mathrm{i}}
\newcommand{\erf}[0]{\text{erf}}
\newcommand{\erfc}[0]{\text{erfc}}

\begin{document}

\title{Cumulant generating functions of a tracer\\ in quenched dense symmetric exclusion processes}

\author{Alexis Poncet}
\affiliation{Sorbonne Universit\'e, CNRS, Laboratoire de Physique Th\'eorique de la Mati\`ere Condens\'ee (LPTMC), 4 Place Jussieu, 75005 Paris, France}
\author{Olivier B\'enichou}
\affiliation{Sorbonne Universit\'e, CNRS, Laboratoire de Physique Th\'eorique de la Mati\`ere Condens\'ee (LPTMC), 4 Place Jussieu, 75005 Paris, France}
\author{Pierre Illien}
\affiliation{Sorbonne Universit\'e, CNRS, Laboratoire de Physico-Chimie des \'Electrolytes et Nanosyst\`emes Interfaciaux (PHENIX), 4 Place Jussieu, 75005 Paris, France}
\date{\today}

\begin{abstract}

The Symmetric Exclusion Process (SEP), where particles hop on a 1D lattice with the restriction that there can only be one particle per site, is a paradigmatic model of interacting particle systems. Recently, it has been shown that the nature of the initial conditions -- annealed or quenched -- has a quantitative impact on the long-time properties of tracer diffusion. However, so far, all the studies in the quenched case focused on the low-density limit of the SEP. Here, we derive the cumulant generating function of the tracer position in the dense limit with quenched initial conditions. Importantly, our approach also allows us to consider the nonequilibrium situations of (i)~a biased tracer in the SEP and (ii) a symmetric tracer in a step of density. In the former situation, we show that the initial conditions have a striking impact, and change the very dependence of the cumulants on the bias.

\end{abstract}

\maketitle

Diffusion of interacting particles under strong confinement gives rise to anomalous subdiffusive behaviours. This is exemplified by single-file diffusion in narrow channels, in which particles cannot bypass each other and remain in the same order. For any tagged particle in the system, this typically leads to a sublinear growth of the fluctuations of positions $\langle X_t^2\rangle \propto\sqrt{t}$ \cite{Harris1965}, in contrast with normal, linear-in-time fluctuations in the absence of confinement or in the absence of interactions. Such a subdiffusive behaviour was observed in different contexts, for instance in porous media or in confined colloidal suspensions \cite{Gupta1995,Hahn1996,Wei2000,Meersmann2000,Lin2005}.

The Symmetric Exclusion Process (SEP) is a classical model of single-file diffusion. In this minimal representation, a one-dimensional lattice is populated by particles at density $\rho$. Each of them performs a symmetric, continuous-time random walk, with the restriction that there can only be one particle per site, which represents the hardcore interactions between particles. It was established that the variance of the position of a tracer scales as $\sqrt{t}$ in the long-time limit, and the prefactor was determined exactly as a function of the density: $\langle X_t^2\rangle \underset{t\to\infty}{\sim} \frac{1-\rho}{\rho}\sqrt{\frac{2t}{\pi}}$ \cite{Arratia1983}. Important developments have been considered during the past decades \cite{Derrida2007,Mallick2015}.

First, even if it has been known for long that  the rescaled position  satisfies a central limit theorem and converges to a fractional Brownian motion with Hurst index $1/4$ \cite{Harris1965,Spitzer1970,Arratia1983,Spohn1990,Peligrad2008}, the large time expression of the higher-order cumulants has been obtained only recently. They have been first  derived in the dense limit $\rho\to1$  \cite{Illien2013a} and dilute limit $\rho\to0$ \cite{Krapivsky2015,Krapivsky2014,Hegde2014} \footnote{In this limit, the SEP is equivalent to the model of hard Brownian particles on a line.}.
A real breakthrough came in 2017 when Imamura, Sasamoto and Mallick derived the full probability law at any density \cite{Imamura2017}.

\begin{figure}[b]
	\centering
	\includegraphics[scale=1]{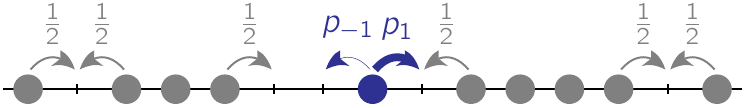}
	\caption{Biased tracer in a SEP. The bath particles (gray) perform symmetric random walks with exclusion. The random walk of the tracer (blue) is biased. Note that, when $p_1=p_{-1}=1/2$, this model is identical to the classical SEP.}
	\label{system}
\end{figure}

Second, while the SEP in its original formulation provides a model of subdiffusion in crowded equilibrium systems, an important extension to non-equilibrium situations has been proposed by considering the general case of a driven tracer in a bath of unbiased random walks (still with exclusion, Fig. \ref{system}). The mean position \cite{Burlatsky1996,Landim1998} and all higher-order moments in the dense limit~\cite{Illien2013a} have been calculated, and shown to grow anomalously as $\sqrt{t}$. Recent extensions of this model concern the case of several driven tracers \cite{Poncet2019} or of a finite system \cite{Lobaskin2020, Ayyer}. Note that a similar behaviour of the first two cumulants is found for a symmetric tracer in an inhomogeneous bath, namely a step of density \cite{Imamura2017}.

Last, recent studies have investigated the sensitivity of tracer diffusion in single-file systems to the initial conditions. Indeed, in analogy with the physics of disordered systems, two distinct situations need to be considered. In the \emph{annealed} case (implicitly assumed in the results reminded above), particles are initially distributed according to an equilibrium state of density $\rho$. In the \emph{quenched} case, the initial positions of the particles are fixed, with the constraint that, at a macroscopic scale, they correspond to a uniform density $\rho$. It was shown that, even if the scaling at large time is the same, the prefactor of the variance  is different for the two settings \cite{Leibovich2013}. This came as a surprise, since  the memory of initial conditions could naively been expected to be lost  at long time (see however \cite{Krug1997} for the description of a similar effect in a different context). Extensions to the calculation of the fourth cumulant, all cumulants and two-time correlation functions in the dilute limit \cite{Sadhua,Krapivsky2015} or to other single-file models  \cite{Cividini2016c,Ooshida2018} have recently been proposed.

However, in the quenched setting, (i) the cumulant generating function (CGF) of a tracer in the SEP has been calculated only in the the dilute limite $\rho\to0$ \footnote{Note, however, that the very specific case of $\rho=1/2$ has been derived using a mapping to the current in the SEP \cite{Derrida2009a, Derrida2009, Sadhua}.}; (ii) nonequilibrium situations, involving a driven tracer or a step of density, have not been investigated in spite of their importance.

In this Letter, we derive the full cumulant generating function of the tracer position  in the dense limit of the SEP with quenched initial conditions. Our approach also allows us to consider two typical nonequilibrium situations: (i) a biased tracer in the SEP and (ii) a symmetric tracer in a step of density (see \cite{SM} for the results corresponding to a combination of both). Strikingly, we show that, in the former situation, the impact of the initial conditions is not only quantitative but also qualitative, and their nature changes the very dependence of the cumulants on the bias.

\emph{Model.---} The system that we study is a biased tracer in the SEP (Fig.~\ref{system}). Particles are initially positioned uniformly at random on the infinite discrete line with a density $\rho$. Each particle has an exponential clock of average $1$ and when the clock ticks, the particle chooses to jump either to the left (with probability $1/2$) or to the right (with probability $1/2$). If the arrival site is empty, the jump is performed. Otherwise, it is canceled. 

One of the particles is assumed to be a tracer and to have different jumping rates: $p_1$ to the right and $p_{-1}$ to the left. The tracer is initially at the origin $X(t=0) = 0$
and we study its displacement with time $X(t)$.
We define the CGF $\psi^{(t)}(k) \equiv \log \langle \ex{\ii kX(t)}\rangle$, whose expansion yields the cumulants $\kappa_n(t)$, defined as $\psi^{(t)}(k) \equiv \sum_{n=1}^\infty \frac{(ik)^n}{n!} \kappa_n(t)$, where $\kappa_1$ is the average displacement and
$\kappa_2$ is the variance. Our goal is the determination, in the quenched setting (see below for a precise definition), of the CGF and the cumulants in the high-density limit $\rho\to 1$. 

\textit{CGF in the high-density limit.}---  Let us first consider a 1D lattice of finite size $N$ in which all the sites are occupied except $M$ of them. We call these empty sites \emph{vacancies}, and their fraction is $ M/N = 1-\rho$. The high density limit of the SEP corresponds to $\rho \to 1$. Instead of looking at the motion of the particles, one can equivalently study the motion of the vacancies, which perform random walks on the line. For simplicity, we adopt here a discrete-time description: at each time step, each vacancy moves to a neighboring site. We will only derive results in the long-time limit, in which this description becomes equivalent to a continuous-time description.

When a vacancy crosses the tracer from left to right, the tracer moves to the left and vice-versa. We number the vacancies and call $Y^j(t)$ the displacement of the tracer generated by the $j$-th vacancy. We have
$X(t) = Y^1(t) + \dots Y^M(t)$. 
The initial positions of the vacancies are called $Z_j$.
$P^{(t)}(X|\{Z_j\})$
is the probability of a displacement $X$ at time $t$ knowing the initial positions of the vacancies. Similarly, $\mathcal{P}^{(t)} (\{Y^j\}|\{Z_j\})$ is the probability that up to time $t$ vacancies induced displacements $\{Y^j\}$ of the tracer.
By definition,
\begin{equation} \label{eq:dense_linkP}
P^{(t)} (X|\{Z_j\}) =
\sum_{Y_1,\dots, Y_M}
\delta_{X, Y_1+\dots+Y_M}
\mathcal{P}^{(t)} (\{Y^j\}|\{Z_j\}).
\end{equation}

In the high density limit ($ M/N\to 0$), the vacancies perform independent random walks and interact independently with the tracer. We neglect events of order $\mathcal{O}[(1-\rho)^2]$ in which two vacancies interact with each other, compared to events of order $\mathcal{O}(1-\rho)$ in which one vacancy interacts with the tracer. This gives exact results in the limit $\rho\to 1$ \cite{Brummelhuis1989a,Brummelhuis1988}. We call $p^{(t)}_Z(Y)$ the probability that in a system with a single vacancy initially at $Z$, the tracer has displacement $Y$ at time $t$. We have $\mathcal{P}^{(t)} (\{Y^j\}|\{Z_j\})  \underset{\rho \to 1}{\sim} \prod_{j=1}^M p^{(t)}_{Z_j}(Y^j)$. Note that there are only two values of $Y$ for which $p^{(t)}_{Z}(Y)$ is non-zero ($Y = 0$ and $\pm 1$ for $Z\lessgtr 0)$).

Using Eq.~\eqref{eq:dense_linkP} and defining the Fourier transform of any site-dependent function as
$\tilde f(k) = \sum_{X=-\infty}^\infty \ex{\ii kX}f(X)$,
we obtain
\begin{equation} \label{eq:1tp_link_prop}
\tilde P^{(t)} (k|\{Z_j\}) \underset{\rho \to 1}{\sim}\prod_{j=1}^M \tilde p^{(t)}_{Z_j}(k).
\end{equation}

For self-consistency, we first consider the case in which the vacancies have equal probability to be on any site (except the origin). This is known in the literature as annealed initial conditions. The cumulant-generating function $\Psi_A^{(t)}(k)$  of $X(t)$ is the logarithm of the average of $\tilde P^{(t)} (k|\{Z_j\})$ on all the initial positions of the vacancies
\begin{align}
 \psi_A^{(t)}(k) &= \log \tilde P_A^{(t)} (k),\\
 \tilde P_A^{(t)} (k) &\equiv \frac{1}{(N-1)^M} \sum_{Z_1, \dots, Z_M\neq 0} \tilde P^{(t)} (k|\{Z_j\}).
\end{align}
Using Eq. \eqref{eq:1tp_link_prop}, we have,  in the limit $\rho \to 1$, $\tilde P_A^{(t)} (k) = \left[1 +\frac{1}{N-1} \sum_{Z\neq 0} \left(\tilde p^{(t)}_{Z}(k)-1\right)\right]^M$, which, in the thermodynamic limit $M, N\to\infty$ with $M/N = 1-\rho$ constant, leads to
\begin{equation} \label{eq:1tp_expr_psi}
\lim_{\rho \to 1} \frac{\psi_A^{(t)}(k)}{1-\rho} = \sum_{Z\neq 0} \left(\tilde p^{(t)}_{Z}(k)-1\right).
\end{equation}

We now turn to the case of quenched initial conditions. The initial positions of the particles are fixed and one averages over multiple realizations of the evolution of the system.
The cumulant-generating function conditioned on the initial
positions of the vacancies is given by \cite{Sadhua}
 \begin{align}
\psi_Q(k, t |\{Z_j\}) &\equiv \log \tilde P^{(t)}(k|\{Z_j\}).
\end{align}
The quenched cumulant-generating function $\psi_Q$ is then defined as the average of this quantity over the initial positions
\begin{align}
 \psi_Q(k, t) &\equiv \frac{1}{(N-1)^M} \sum_{Z_1, \dots Z_M} \psi_Q(k, t |\{Z_j\}).
 \label{psiQaverage}
\end{align}
Using Eq.~\eqref{eq:1tp_link_prop}, and taking the thermodynamic limit, we find that the high-density limit of the quenched cumulant-generating function reads
\begin{equation} \label{eq:1tp_psiQ}
\lim_{\rho \to 1} \frac{\psi_Q(k, t)}{1-\rho}=
\sum_{Z\neq 0} \log \tilde p^{(t)}_{Z}(k).
\end{equation}
Let us emphasize the difference between Eqs. \eqref{eq:1tp_expr_psi} and \eqref{eq:1tp_psiQ}: in the annealed case, the CGF is a linear combination of the single-vacancy propagators, as opposed to the nonlinear dependence of the quenched case. This structure is reminiscent of the expressions obtained in the opposite limit of a dilute SEP \cite{Sadhua}.

Explicit results can be obtained by noting that the generating function associated with the single vacancy propagator, defined as $\hat{\tilde{p}}_{Z}(k,\xi)= \sum_{t=0}^\infty  p^{(t)}_{Z}(k) \xi^t$, has been determined in the calculation of the annealed CGF and reads \cite{Illien2013a}
\begin{equation}
\label{singlevacprop_Laplace}
\hat{\tilde{p}}_{Z}(k,\xi)= \frac{1}{1-\xi} \left[  1+(\ex{\ii \mu k}-1) \frac{1-\hat{f}_{-\mu}(\xi)}{1-\hat{f}_{1}(\xi)\hat{f}_{-1}(\xi)}\hat{f}_Z(\xi) \right]
\end{equation}
where $\mu \equiv \text{sgn}(Z)$ and $f_Z^{(t)}$ is the probability for a vacancy to reach the origin for the first time at $t$ starting from $Z$. Its generating function reads
\begin{equation}
\label{ }
\hat{f}_Z(\xi)=\frac{1+\mu s}{1+\mu s \alpha}\alpha^{|Z|}
\end{equation}
with $\alpha=(1-\sqrt{1-\xi^2})/\xi$ and $s=p_1-p_{-1}$.

\begin{figure*}
	\centering 
	\includegraphics[scale=1]{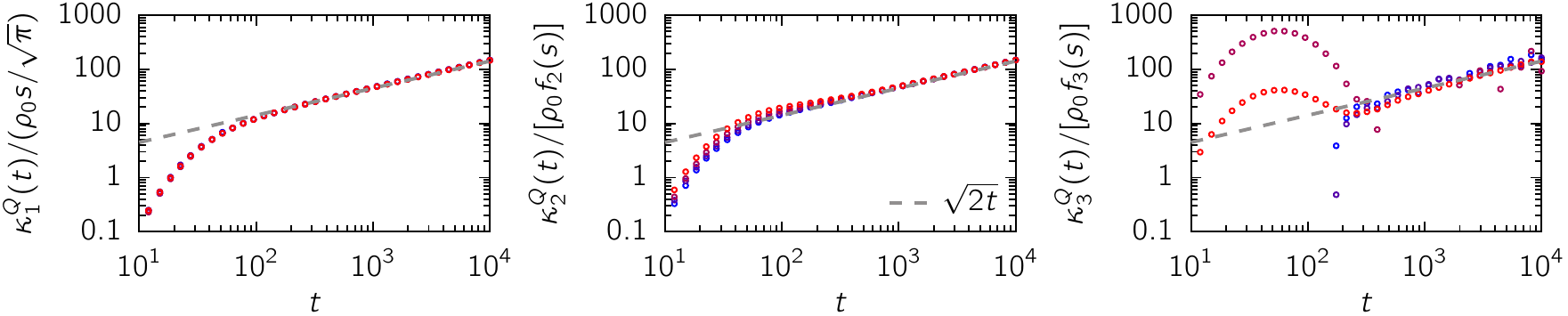}
	\caption{Cumulants $\kappa_1^Q$, $\kappa_2^Q$, $\kappa_3^Q$ of a biased tracer in the quenched SEP at density $\rho = 0.95$ for biases $s=p_1-p_{-1}=0.2$, 0.5, 0.8 and 1 (blue to red). The numerical simulations (circles) are performed with deterministic initial conditions.
	We denote $\rho_0 = 1-\rho$ and
	we compute the scalings for $\kappa_2^Q$ and $\kappa_3^Q$ from Eq.~\eqref{eq:1tp_resPsiQ}: $f_2(s) = \left[1 + s^2(1-\sqrt 2)\right]/\sqrt{2\pi}$ and
	$f_3(s) = \left[6\sqrt{2}(3+s^2)\arctan(2^{-1/2})/\pi
	-1 - 3\sqrt{2} - 3(\sqrt{2} - 1)s^2\right]/\sqrt{4\pi}$.
	In the three sub-figures, the dashed gray line corresponds to our prediction: $\sqrt{2t}$.
		}
	\label{biased}
\end{figure*}

\emph{Large time limit of the quenched CGF.---}
Taking the scaling limit $\xi\to 1$ with $(1-\xi)|Z|^2$ kept constant (corresponding to the scaling limit $t~\to~\infty$ with $|Z|/\sqrt{t}$ constant), Eq.~\eqref{singlevacprop_Laplace} leads, after inverse Laplace transform, to
\begin{equation}
\tilde p^{(t)}_{Z}(k) \underset{t\to\infty}{\sim}
1 + p_\mu \left(\ex{\ii\mu k} -1\right)
\text{erfc}\left(\frac{|Z|}{\sqrt{2t}}\right).
\label{psingle_real}
\end{equation}
Using this result in Eq.~\eqref{eq:1tp_psiQ}, the large-time limit of the involved Riemann sum yields 
\begin{widetext}
\begin{equation}
\label{eq:1tp_resPsiQ}
\lim_{\rho \to 1} \frac{\psi_Q(k, t)}{1-\rho} \underset{t\to\infty}{\sim}
\sqrt{2t} \int_0^\infty \dd z \,   \log \Big\{\left[1 + p_1 \left(\ex{\ii k} -1\right)
\text{erfc} z\right]
\left[1 + p_{-1} \left(\ex{-\ii k} -1\right)
\text{erfc} z\right]\Big\}.
\end{equation}
\end{widetext}
This cumulant generating function, derived for a biased tracer in a high-density SEP, is the key result of our Letter.

\textit{Symmetric tracer.}--- We first focus on the situation where the tracer is symmetric, i.e. $p_1=p_{-1}=1/2$. The CGF [Eq.~\eqref{eq:1tp_resPsiQ}] takes the simple form:
\begin{align}
&\lim_{\rho \to 1} \frac{\psi_Q(k, t)}{1-\rho}\nonumber\\
& \underset{t\to\infty}{\sim}
\sqrt{2t} \int_0^\infty dz\log\left[1 - \sin^2\left(\frac{k}{2}\right) \text{erfc}( z) \text{erfc}(-z)\right].
\label{CGF_sym_highdens}
\end{align}
The known cumulants $\kappa_2^Q$ and $\kappa_4^Q$ are retrieved \cite{Krapivsky2015}, but more generally all the cumulants can be deduced from Eq. \eqref{CGF_sym_highdens} by a simple series expansion.

\begin{figure*}
	\centering
	\includegraphics[scale=1]{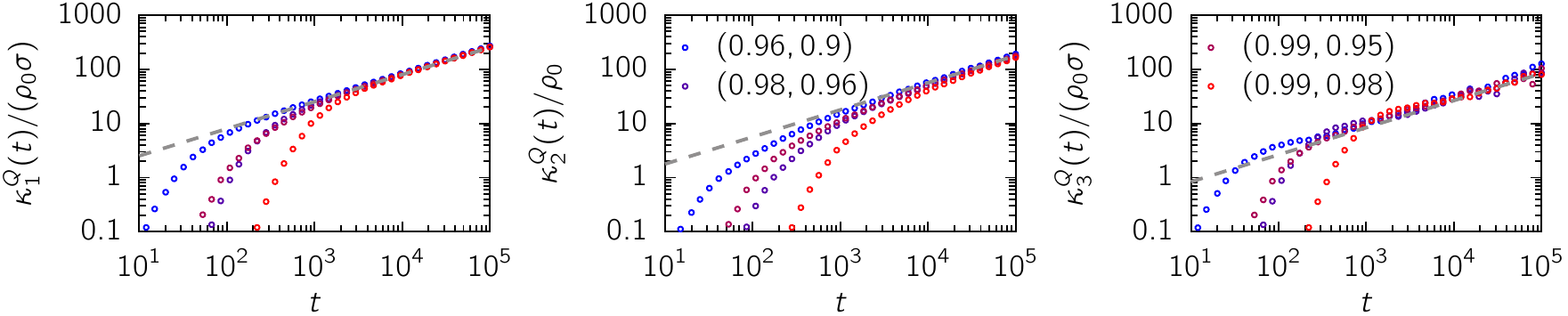}
	\caption{Cumulants $\kappa_1^Q$, $\kappa_2^Q$, $\kappa_3^Q$ of a tracer in the quenched SEP with densities $\rho_-$ behind the tracer and $\rho_+$ in front of it.
		From blue to red, $(\rho_-, \rho_+) = (0.96, 0.9), (0.98, 0.96), (0.99, 0.95), (0.99, 0.98)$. The numerical simulations (cicles) are performed with deterministic initial conditions. The gray lines are the predictions deduced from Eq.~\eqref{eq:1tp_resPsiQ_step}. We denote $\rho_0 = 1-\rho$.}
	\label{step}
\end{figure*}

\textit{Biased tracer.}--- In the case of a biased tracer ($s=p_1-p_{-1}\neq 0$), all the cumulants can be computed from Eq. \eqref{eq:1tp_resPsiQ}, and the first three read
\begin{align}
&\lim_{\rho \to 1} \frac{\kappa_1^Q(t)}{1-\rho} \underset{t\to\infty}{\sim} s \sqrt\frac{2t}{\pi} \label{eq:1tp_resK1Q}, \\
&\lim_{\rho \to 1} \frac{\kappa_2^Q(t)}{1-\rho} \underset{t\to\infty}{\sim} \sqrt\frac{t}{\pi}
\left(1 + s^2(1-\sqrt{2})\right) \label{eq:1tp_resK2Q}, \\
&\lim_{\rho \to 1} \frac{\kappa_3^Q(t)}{1-\rho} \underset{t\to\infty}{\sim} s\sqrt\frac{t}{2\pi} \nonumber \\
&\times \left[\frac{6\sqrt{2}(3+s^2)}{\pi}\arctan\frac{1}{\sqrt{2}}
 -1 - 3\sqrt{2} - 3(\sqrt{2} - 1)s^2\right] \label{eq:1tp_resK3Q}.
\end{align}
Several comments are in order: (i) the first cumulant is identical in the quenched and in the annealed settings \cite{Illien2013a}, as in the opposite limit of low-density (in which only the first cumulant was computed \cite{Leibovich2013}). (ii) The identity between odd cumulants on the one hand and even cumulants on the other hand, shown in the annealed case \cite{Illien2013a}, does not hold for quenched initial conditions.  (iii) Importantly, there is a strong impact of the initial conditions on the second cumulant. Indeed, in the annealed case, it was shown that, the variance is independent of the bias, and reads $\underset{\rho \to 1} {\lim}\frac{\kappa_2^A (t)}{1-\rho}\underset{t\to\infty}{\sim}   \sqrt{\frac{2t}{\pi}}$, as opposed with Eq. \eqref{eq:1tp_resK2Q}. This very striking difference emphasizes the importance of the initial conditions of the system on the long-time dynamics of this nonequilibrium situation, with both qualitative and quantitative consequences. Agreement with numerical simulations  performed with the choice of deterministic initial conditions \cite{SM} is displayed on Fig.~\ref{biased}.

\textit{Step of density.}--- Finally, we consider an unbiased tracer in a step of density, as studied in the annealed case \cite{Gleb,Imamura2017}. The density in front of the tracer (resp. behind the tracer) is denoted $\rho_+$ (resp. $\rho_-$). The average density is $\rho = (\rho_+ + \rho_-) / 2$ and
we define the step $\sigma = (\rho_- - \rho_+)/[2(1-\rho)]$.
 For annealed initial conditions in the high-density limit, it is shown in Supplemental Material \cite{SM}, and in agreement with the general results of \cite{Imamura2017}, that all odd (resp. even) cumulants are identical and read
 \begin{align}
\lim_{\rho \to 1} \frac{\kappa_\text{odd}^A(t)}{1-\rho} &\underset{t\to\infty}{\sim} \sigma \sqrt\frac{2t}{\pi},
\label{K_odd_step_ann} \\
\lim_{\rho \to 1} \frac{\kappa_\text{even}^A(t)}{1-\rho} &\underset{t\to\infty}{\sim}  \sqrt\frac{2t}{\pi},
\label{K_even_step_ann}
\end{align}
Note that the cumulants have the same expression than for a biased tracer in a homogeneous bath of density $\rho$ if the substution $s = p_{1} - p_{-1} \mapsto \sigma$ is made \cite{Illien2013a}.

For quenched initial conditions, the previous derivation can be adapted \cite{SM}, and leads to
\begin{multline}
\label{eq:1tp_resPsiQ_step}
\lim_{\rho \to 1} \frac{\psi_Q(k, t)}{1-\rho}\underset{t\to\infty}{\sim}\sqrt{2t} \\  
 \times\sum_{\mu=\pm 1} (1+\mu \sigma)\int_0^\infty \dd z \, \log\left[1 + \frac{1}{2} \left(\ex{\ii \mu k} -1\right)
\text{erfc} z\right].
\end{multline}
One striking consequence is that the even cumulants are the same as those for a tracer in a homogeneous bath whose effective density is the average of $\rho_+$ and $\rho_-$, as deduced from Eq. \eqref{CGF_sym_highdens}.

The odd cumulants are all proportional to the step $\sigma$, the lowest ones read
\begin{align}
\lim_{\rho \to 1} \frac{\kappa_1^Q(t)}{1-\rho} &\underset{t\to\infty}{\sim} \sigma \sqrt\frac{2t}{\pi},
\label{K1_step} \\
\lim_{\rho \to 1} \frac{\kappa_3^Q(t)}{1-\rho} &\underset{t\to\infty}{\sim}
\sigma \sqrt\frac{t}{2\pi}\left[\frac{6\sqrt{2}}{\pi}\arctan\frac{1}{\sqrt{2}}-1\right].
\label{K3_step}
\end{align}
We note that while the average displacement $\kappa_1^Q$ is identical to the case of annealed initial conditions [Eq. \eqref{K_odd_step_ann}], the prefactors of the higher-order odd cumulants are modified. Last, we compare in Fig. ~\ref{step} our analytical prediction against the numerical simulations and again observe a very good agreement. Finally, we emphasize that the CGF can be derived in the general case combining a biased tracer in a density step \cite{SM}.\\

 As a last observation, we note the strong similarity between the CGF given by Eq. \eqref{CGF_sym_highdens} in the dense regime $\rho\to1$ and the CGF derived in the opposite regime of $\rho \to 0$ \cite{Krapivsky2015,Sadhua}. The functional dependence of the CGF on $\rho$ in these two regimes could lead us to propose the following expression for the CGF at arbitrary density:
\begin{align}
&\psi^\text{approx}_Q(k, t)\underset{t\to\infty}{\sim}\rho(1-\rho)\sqrt{2t} \nonumber \\
&\times \int_0^\infty dz\log\left[1 - \sin^2\left(\frac{k}{2\rho}\right) \text{erfc}( z) \text{erfc}(-z)\right].
\label{CGF_sym_guess}
\end{align}
The quantitative agreement between the first cumulants computed from Eq. \eqref{CGF_sym_guess}  and numerical simulations \cite{SM} could further point towards the exact nature of this expression. However, even though the second cumulant  that could be deduced from \eqref{CGF_sym_guess} is identical to the exact expression obtained from macroscopic fluctuation theory \cite{Krapivsky2015}, this is not true for the fourth cumulant. Finally, Eq. \eqref{CGF_sym_guess} is a good approximation for arbitrary $\rho$ but is exact only in the limits of $\rho\to0$ and $\rho\to1$. The calculation of the CGF with quenched initial conditions at arbitrary density, that would be the counterpart of the result obtained in the annealed case \cite{Imamura2017}, remains a challenging open question.

\bibliographystyle{apsrev4-1}


\begin{thebibliography}{35}%
\makeatletter
\providecommand \@ifxundefined [1]{%
 \@ifx{#1\undefined}
}%
\providecommand \@ifnum [1]{%
 \ifnum #1\expandafter \@firstoftwo
 \else \expandafter \@secondoftwo
 \fi
}%
\providecommand \@ifx [1]{%
 \ifx #1\expandafter \@firstoftwo
 \else \expandafter \@secondoftwo
 \fi
}%
\providecommand \natexlab [1]{#1}%
\providecommand \enquote  [1]{``#1''}%
\providecommand \bibnamefont  [1]{#1}%
\providecommand \bibfnamefont [1]{#1}%
\providecommand \citenamefont [1]{#1}%
\providecommand \href@noop [0]{\@secondoftwo}%
\providecommand \href [0]{\begingroup \@sanitize@url \@href}%
\providecommand \@href[1]{\@@startlink{#1}\@@href}%
\providecommand \@@href[1]{\endgroup#1\@@endlink}%
\providecommand \@sanitize@url [0]{\catcode `\\12\catcode `\$12\catcode
  `\&12\catcode `\#12\catcode `\^12\catcode `\_12\catcode `\%12\relax}%
\providecommand \@@startlink[1]{}%
\providecommand \@@endlink[0]{}%
\providecommand \url  [0]{\begingroup\@sanitize@url \@url }%
\providecommand \@url [1]{\endgroup\@href {#1}{\urlprefix }}%
\providecommand \urlprefix  [0]{URL }%
\providecommand \Eprint [0]{\href }%
\providecommand \doibase [0]{http://dx.doi.org/}%
\providecommand \selectlanguage [0]{\@gobble}%
\providecommand \bibinfo  [0]{\@secondoftwo}%
\providecommand \bibfield  [0]{\@secondoftwo}%
\providecommand \translation [1]{[#1]}%
\providecommand \BibitemOpen [0]{}%
\providecommand \bibitemStop [0]{}%
\providecommand \bibitemNoStop [0]{.\EOS\space}%
\providecommand \EOS [0]{\spacefactor3000\relax}%
\providecommand \BibitemShut  [1]{\csname bibitem#1\endcsname}%
\let\auto@bib@innerbib\@empty
\bibitem [{\citenamefont {Harris}(1965)}]{Harris1965}%
  \BibitemOpen
  \bibfield  {author} {\bibinfo {author} {\bibfnamefont {T.}~\bibnamefont
  {Harris}},\ }\href@noop {} {\bibfield  {journal} {\bibinfo  {journal}
  {Journal of Applied Probability}\ }\textbf {\bibinfo {volume} {2}},\ \bibinfo
  {pages} {323} (\bibinfo {year} {1965})}\BibitemShut {NoStop}%
\bibitem [{\citenamefont {Gupta}\ \emph {et~al.}(1995)\citenamefont {Gupta},
  \citenamefont {Nivarthi}, \citenamefont {McCormick},\ and\ \citenamefont
  {{Ted Davis}}}]{Gupta1995}%
  \BibitemOpen
  \bibfield  {author} {\bibinfo {author} {\bibfnamefont {V.}~\bibnamefont
  {Gupta}}, \bibinfo {author} {\bibfnamefont {S.~S.}\ \bibnamefont {Nivarthi}},
  \bibinfo {author} {\bibfnamefont {A.~V.}\ \bibnamefont {McCormick}}, \ and\
  \bibinfo {author} {\bibfnamefont {H.}~\bibnamefont {{Ted Davis}}},\ }\href
  {http://ac.els-cdn.com/S000926149501246X/1-s2.0-
  S000926149501246X-main.pdf?{\_}tid=47f35abc-94fb-11e4-90e7- 00000aacb35f{\&
  }acdnat=1420476949{\_}6e51a7fede5c322b97d633d789588a4f http:/
  /www.sciencedirect.com/science/article/pii/ S000926149501246X} {\bibfield
  {journal} {\bibinfo  {journal} {Chemical Physics Letters}\ }\textbf {\bibinfo
  {volume} {247}},\ \bibinfo {pages} {596} (\bibinfo {year}
  {1995})}\BibitemShut {NoStop}%
\bibitem [{\citenamefont {Hahn}\ \emph {et~al.}(1996)\citenamefont {Hahn},
  \citenamefont {K{\"{a}}rger},\ and\ \citenamefont {Kukla}}]{Hahn1996}%
  \BibitemOpen
  \bibfield  {author} {\bibinfo {author} {\bibfnamefont {K.}~\bibnamefont
  {Hahn}}, \bibinfo {author} {\bibfnamefont {J.}~\bibnamefont {K{\"{a}}rger}},
  \ and\ \bibinfo {author} {\bibfnamefont {V.}~\bibnamefont {Kukla}},\ }\href
  {http://journals.aps.org/prl/pdf/10.1103/ PhysRevLett.76.2762
  http://journals.aps.org/prl/abstract/ 10.1103/PhysRevLett.76.2762} {\bibfield
   {journal} {\bibinfo  {journal} {Physical Review Letters}\ }\textbf {\bibinfo
  {volume} {76}},\ \bibinfo {pages} {2762} (\bibinfo {year}
  {1996})}\BibitemShut {NoStop}%
\bibitem [{\citenamefont {Wei}(2000)}]{Wei2000}%
  \BibitemOpen
  \bibfield  {author} {\bibinfo {author} {\bibfnamefont {Q.}~\bibnamefont
  {Wei}},\ }\href {\doibase 10.1126/science.287.5453.625} {\bibfield  {journal}
  {\bibinfo  {journal} {Science}\ }\textbf {\bibinfo {volume} {287}},\ \bibinfo
  {pages} {625} (\bibinfo {year} {2000})}\BibitemShut {NoStop}%
\bibitem [{\citenamefont {Meersmann}\ \emph {et~al.}(2000)\citenamefont
  {Meersmann}, \citenamefont {Logan}, \citenamefont {Simonutti}, \citenamefont
  {Caldarelli}, \citenamefont {Comotti}, \citenamefont {Sozzani}, \citenamefont
  {Kaiser},\ and\ \citenamefont {Pines}}]{Meersmann2000}%
  \BibitemOpen
  \bibfield  {author} {\bibinfo {author} {\bibfnamefont {T.}~\bibnamefont
  {Meersmann}}, \bibinfo {author} {\bibfnamefont {J.~W.}\ \bibnamefont
  {Logan}}, \bibinfo {author} {\bibfnamefont {R.}~\bibnamefont {Simonutti}},
  \bibinfo {author} {\bibfnamefont {S.}~\bibnamefont {Caldarelli}}, \bibinfo
  {author} {\bibfnamefont {A.}~\bibnamefont {Comotti}}, \bibinfo {author}
  {\bibfnamefont {P.}~\bibnamefont {Sozzani}}, \bibinfo {author} {\bibfnamefont
  {L.~G.}\ \bibnamefont {Kaiser}}, \ and\ \bibinfo {author} {\bibfnamefont
  {A.}~\bibnamefont {Pines}},\ }\href {\doibase 10.1021/jp002322v} {\bibfield
  {journal} {\bibinfo  {journal} {Journal of Physical Chemistry A}\ }\textbf
  {\bibinfo {volume} {104}},\ \bibinfo {pages} {11669} (\bibinfo {year}
  {2000})}\BibitemShut {NoStop}%
\bibitem [{\citenamefont {Lin}\ \emph {et~al.}(2005)\citenamefont {Lin},
  \citenamefont {Meron}, \citenamefont {Cui}, \citenamefont {Rice},\ and\
  \citenamefont {Diamant}}]{Lin2005}%
  \BibitemOpen
  \bibfield  {author} {\bibinfo {author} {\bibfnamefont {B.}~\bibnamefont
  {Lin}}, \bibinfo {author} {\bibfnamefont {M.}~\bibnamefont {Meron}}, \bibinfo
  {author} {\bibfnamefont {B.}~\bibnamefont {Cui}}, \bibinfo {author}
  {\bibfnamefont {S.~A.}\ \bibnamefont {Rice}}, \ and\ \bibinfo {author}
  {\bibfnamefont {H.}~\bibnamefont {Diamant}},\ }\href {\doibase
  10.1103/PhysRevLett.94.216001} {\bibfield  {journal} {\bibinfo  {journal}
  {Physical Review Letters}\ }\textbf {\bibinfo {volume} {94}},\ \bibinfo
  {pages} {216001} (\bibinfo {year} {2005})}\BibitemShut {NoStop}%
\bibitem [{\citenamefont {Arratia}(1983)}]{Arratia1983}%
  \BibitemOpen
  \bibfield  {author} {\bibinfo {author} {\bibfnamefont {R.}~\bibnamefont
  {Arratia}},\ }\href {\doibase 10.1214/aop/1176993602} {\bibfield  {journal}
  {\bibinfo  {journal} {The Annals of Probability}\ }\textbf {\bibinfo {volume}
  {11}},\ \bibinfo {pages} {362} (\bibinfo {year} {1983})}\BibitemShut
  {NoStop}%
\bibitem [{\citenamefont {Derrida}(2007)}]{Derrida2007}%
  \BibitemOpen
  \bibfield  {author} {\bibinfo {author} {\bibfnamefont {B.}~\bibnamefont
  {Derrida}},\ }\href {\doibase 10.1088/1742-5468/2007/07/P07023} {\bibfield
  {journal} {\bibinfo  {journal} {Journal of Statistical Mechanics: Theory and
  Experiment}\ }\textbf {\bibinfo {volume} {2007}},\ \bibinfo {pages} {P07023}
  (\bibinfo {year} {2007})}\BibitemShut {NoStop}%
\bibitem [{\citenamefont {Mallick}(2015)}]{Mallick2015}%
  \BibitemOpen
  \bibfield  {author} {\bibinfo {author} {\bibfnamefont {K.}~\bibnamefont
  {Mallick}},\ }\href {\doibase 10.1016/j.physa.2014.07.046} {\bibfield
  {journal} {\bibinfo  {journal} {Physica A}\ }\textbf {\bibinfo {volume}
  {418}},\ \bibinfo {pages} {17} (\bibinfo {year} {2015})}\BibitemShut
  {NoStop}%
\bibitem [{\citenamefont {Spitzer}(1970)}]{Spitzer1970}%
  \BibitemOpen
  \bibfield  {author} {\bibinfo {author} {\bibfnamefont {F.}~\bibnamefont
  {Spitzer}},\ }\href {\doibase 10.1016/0001-8708(70)90034-4} {\bibfield
  {journal} {\bibinfo  {journal} {Advances in Mathematics}\ }\textbf {\bibinfo
  {volume} {5}},\ \bibinfo {pages} {246} (\bibinfo {year} {1970})}\BibitemShut
  {NoStop}%
\bibitem [{\citenamefont {Spohn}(1990)}]{Spohn1990}%
  \BibitemOpen
  \bibfield  {author} {\bibinfo {author} {\bibfnamefont {H.}~\bibnamefont
  {Spohn}},\ }\href {\doibase 10.1007/BF01334748} {\bibfield  {journal}
  {\bibinfo  {journal} {Journal of Statistical Physics}\ }\textbf {\bibinfo
  {volume} {59}},\ \bibinfo {pages} {1227} (\bibinfo {year}
  {1990})}\BibitemShut {NoStop}%
\bibitem [{\citenamefont {Peligrad}\ and\ \citenamefont
  {Sethuraman}(2008)}]{Peligrad2008}%
  \BibitemOpen
  \bibfield  {author} {\bibinfo {author} {\bibfnamefont {M.}~\bibnamefont
  {Peligrad}}\ and\ \bibinfo {author} {\bibfnamefont {S.}~\bibnamefont
  {Sethuraman}},\ }\href {http://alea.impa.br/articles/v4/04-12.pdf
  http://arxiv.org/abs/0711.0017} {\bibfield  {journal} {\bibinfo  {journal}
  {Alea}\ }\textbf {\bibinfo {volume} {4}},\ \bibinfo {pages} {245} (\bibinfo
  {year} {2008})}\BibitemShut {NoStop}%
\bibitem [{\citenamefont {Illien}\ \emph {et~al.}(2013)\citenamefont {Illien},
  \citenamefont {B{\'{e}}nichou}, \citenamefont {Mej{\'{i}}a-Monasterio},
  \citenamefont {Oshanin},\ and\ \citenamefont {Voituriez}}]{Illien2013a}%
  \BibitemOpen
  \bibfield  {author} {\bibinfo {author} {\bibfnamefont {P.}~\bibnamefont
  {Illien}}, \bibinfo {author} {\bibfnamefont {O.}~\bibnamefont
  {B{\'{e}}nichou}}, \bibinfo {author} {\bibfnamefont {C.}~\bibnamefont
  {Mej{\'{i}}a-Monasterio}}, \bibinfo {author} {\bibfnamefont {G.}~\bibnamefont
  {Oshanin}}, \ and\ \bibinfo {author} {\bibfnamefont {R.}~\bibnamefont
  {Voituriez}},\ }\href@noop {} {\bibfield  {journal} {\bibinfo  {journal}
  {Physical Review Letters}\ }\textbf {\bibinfo {volume} {111}},\ \bibinfo
  {pages} {38102} (\bibinfo {year} {2013})}\BibitemShut {NoStop}%
\bibitem [{\citenamefont {Krapivsky}\ \emph {et~al.}(2015)\citenamefont
  {Krapivsky}, \citenamefont {Mallick},\ and\ \citenamefont
  {Sadhu}}]{Krapivsky2015}%
  \BibitemOpen
  \bibfield  {author} {\bibinfo {author} {\bibfnamefont {P.~L.}\ \bibnamefont
  {Krapivsky}}, \bibinfo {author} {\bibfnamefont {K.}~\bibnamefont {Mallick}},
  \ and\ \bibinfo {author} {\bibfnamefont {T.}~\bibnamefont {Sadhu}},\ }\href
  {\doibase 10.1007/s10955-015-1291-0} {\bibfield  {journal} {\bibinfo
  {journal} {Journal of Statistical Physics}\ }\textbf {\bibinfo {volume}
  {160}},\ \bibinfo {pages} {885} (\bibinfo {year} {2015})}\BibitemShut
  {NoStop}%
\bibitem [{\citenamefont {Krapivsky}\ \emph {et~al.}(2014)\citenamefont
  {Krapivsky}, \citenamefont {Mallick},\ and\ \citenamefont
  {Sadhu}}]{Krapivsky2014}%
  \BibitemOpen
  \bibfield  {author} {\bibinfo {author} {\bibfnamefont {P.~L.}\ \bibnamefont
  {Krapivsky}}, \bibinfo {author} {\bibfnamefont {K.}~\bibnamefont {Mallick}},
  \ and\ \bibinfo {author} {\bibfnamefont {T.}~\bibnamefont {Sadhu}},\ }\href
  {\doibase 10.1103/PhysRevLett.113.078101} {\bibfield  {journal} {\bibinfo
  {journal} {Physical Review Letters}\ }\textbf {\bibinfo {volume} {113}},\
  \bibinfo {pages} {078101} (\bibinfo {year} {2014})}\BibitemShut {NoStop}%
\bibitem [{\citenamefont {Hegde}\ \emph {et~al.}(2014)\citenamefont {Hegde},
  \citenamefont {Sabhapandit},\ and\ \citenamefont {Dhar}}]{Hegde2014}%
  \BibitemOpen
  \bibfield  {author} {\bibinfo {author} {\bibfnamefont {C.}~\bibnamefont
  {Hegde}}, \bibinfo {author} {\bibfnamefont {S.}~\bibnamefont {Sabhapandit}},
  \ and\ \bibinfo {author} {\bibfnamefont {A.}~\bibnamefont {Dhar}},\ }\href
  {\doibase 10.1103/PhysRevLett.113.120601} {\bibfield  {journal} {\bibinfo
  {journal} {Phys. Rev. Lett.}\ }\textbf {\bibinfo {volume} {113}},\ \bibinfo
  {pages} {120601} (\bibinfo {year} {2014})}\BibitemShut {NoStop}%
\bibitem [{Note1()}]{Note1}%
  \BibitemOpen
  \bibinfo {note} {In this limit, the SEP is equivalent to the model of hard
  Brownian particles on a line.}\BibitemShut {Stop}%
\bibitem [{\citenamefont {Imamura}\ \emph {et~al.}(2017)\citenamefont
  {Imamura}, \citenamefont {Mallick},\ and\ \citenamefont
  {Sasamoto}}]{Imamura2017}%
  \BibitemOpen
  \bibfield  {author} {\bibinfo {author} {\bibfnamefont {T.}~\bibnamefont
  {Imamura}}, \bibinfo {author} {\bibfnamefont {K.}~\bibnamefont {Mallick}}, \
  and\ \bibinfo {author} {\bibfnamefont {T.}~\bibnamefont {Sasamoto}},\ }\href
  {\doibase 10.1103/PhysRevLett.118.160601} {\bibfield  {journal} {\bibinfo
  {journal} {Phys. Rev. Lett.}\ }\textbf {\bibinfo {volume} {118}},\ \bibinfo
  {pages} {160601} (\bibinfo {year} {2017})}\BibitemShut {NoStop}%
\bibitem [{\citenamefont {Burlatsky}\ \emph {et~al.}(1996)\citenamefont
  {Burlatsky}, \citenamefont {Oshanin}, \citenamefont {Moreau},\ and\
  \citenamefont {Reinhardt}}]{Burlatsky1996}%
  \BibitemOpen
  \bibfield  {author} {\bibinfo {author} {\bibfnamefont {S.~F.}\ \bibnamefont
  {Burlatsky}}, \bibinfo {author} {\bibfnamefont {G.}~\bibnamefont {Oshanin}},
  \bibinfo {author} {\bibfnamefont {M.}~\bibnamefont {Moreau}}, \ and\ \bibinfo
  {author} {\bibfnamefont {W.~P.}\ \bibnamefont {Reinhardt}},\ }\href@noop {}
  {\bibfield  {journal} {\bibinfo  {journal} {Phys. Rev. E}\ }\textbf {\bibinfo
  {volume} {54}},\ \bibinfo {pages} {3165} (\bibinfo {year}
  {1996})}\BibitemShut {NoStop}%
\bibitem [{\citenamefont {Landim}\ \emph {et~al.}(1998)\citenamefont {Landim},
  \citenamefont {Olla},\ and\ \citenamefont {Volchan}}]{Landim1998}%
  \BibitemOpen
  \bibfield  {author} {\bibinfo {author} {\bibfnamefont {C.}~\bibnamefont
  {Landim}}, \bibinfo {author} {\bibfnamefont {S.}~\bibnamefont {Olla}}, \ and\
  \bibinfo {author} {\bibfnamefont {S.~B.}\ \bibnamefont {Volchan}},\ }\href
  {\doibase 10.1007/s002200050300} {\bibfield  {journal} {\bibinfo  {journal}
  {Communications in Mathematical Physics}\ }\textbf {\bibinfo {volume}
  {192}},\ \bibinfo {pages} {287} (\bibinfo {year} {1998})}\BibitemShut
  {NoStop}%
\bibitem [{\citenamefont {Poncet}\ \emph {et~al.}(2019)\citenamefont {Poncet},
  \citenamefont {B{\'{e}}nichou}, \citenamefont {D{\'{e}}mery},\ and\
  \citenamefont {Oshanin}}]{Poncet2019}%
  \BibitemOpen
  \bibfield  {author} {\bibinfo {author} {\bibfnamefont {A.}~\bibnamefont
  {Poncet}}, \bibinfo {author} {\bibfnamefont {O.}~\bibnamefont
  {B{\'{e}}nichou}}, \bibinfo {author} {\bibfnamefont {V.}~\bibnamefont
  {D{\'{e}}mery}}, \ and\ \bibinfo {author} {\bibfnamefont {G.}~\bibnamefont
  {Oshanin}},\ }\href {\doibase 10.1103/physrevresearch.1.033089} {\bibfield
  {journal} {\bibinfo  {journal} {Phys. Rev. Research}\ }\textbf {\bibinfo
  {volume} {1}},\ \bibinfo {pages} {033089} (\bibinfo {year}
  {2019})}\BibitemShut {NoStop}%
\bibitem [{\citenamefont {Lobaskin}\ and\ \citenamefont
  {Evans}(2020)}]{Lobaskin2020}%
  \BibitemOpen
  \bibfield  {author} {\bibinfo {author} {\bibfnamefont {I.}~\bibnamefont
  {Lobaskin}}\ and\ \bibinfo {author} {\bibfnamefont {M.~R.}\ \bibnamefont
  {Evans}},\ }\href@noop {} {\bibfield  {journal} {\bibinfo  {journal} {J.
  Stat. Mech.}\ ,\ \bibinfo {pages} {053202}} (\bibinfo {year}
  {2020})}\BibitemShut {NoStop}%
\bibitem [{\citenamefont {Ayyer}()}]{Ayyer}%
  \BibitemOpen
  \bibfield  {author} {\bibinfo {author} {\bibfnamefont {A.}~\bibnamefont
  {Ayyer}},\ }\href@noop {} {\ }\Eprint {http://arxiv.org/abs/2001.02425}
  {arXiv:2001.02425} \BibitemShut {NoStop}%
\bibitem [{\citenamefont {Leibovich}\ and\ \citenamefont
  {Barkai}(2013)}]{Leibovich2013}%
  \BibitemOpen
  \bibfield  {author} {\bibinfo {author} {\bibfnamefont {N.}~\bibnamefont
  {Leibovich}}\ and\ \bibinfo {author} {\bibfnamefont {E.}~\bibnamefont
  {Barkai}},\ }\href {\doibase 10.1103/PhysRevE.88.032107} {\bibfield
  {journal} {\bibinfo  {journal} {Physical Review E}\ }\textbf {\bibinfo
  {volume} {88}},\ \bibinfo {pages} {032107} (\bibinfo {year}
  {2013})}\BibitemShut {NoStop}%
\bibitem [{\citenamefont {Krug}\ \emph {et~al.}(1997)\citenamefont {Krug},
  \citenamefont {Kallabis}, \citenamefont {Majumdar}, \citenamefont {Cornell},
  \citenamefont {Bray},\ and\ \citenamefont {Sire}}]{Krug1997}%
  \BibitemOpen
  \bibfield  {author} {\bibinfo {author} {\bibfnamefont {J.}~\bibnamefont
  {Krug}}, \bibinfo {author} {\bibfnamefont {H.}~\bibnamefont {Kallabis}},
  \bibinfo {author} {\bibfnamefont {S.}~\bibnamefont {Majumdar}}, \bibinfo
  {author} {\bibfnamefont {S.}~\bibnamefont {Cornell}}, \bibinfo {author}
  {\bibfnamefont {a.}~\bibnamefont {Bray}}, \ and\ \bibinfo {author}
  {\bibfnamefont {C.}~\bibnamefont {Sire}},\ }\href {\doibase
  10.1103/PhysRevE.56.2702} {\bibfield  {journal} {\bibinfo  {journal}
  {Physical Review E}\ }\textbf {\bibinfo {volume} {56}},\ \bibinfo {pages}
  {2702} (\bibinfo {year} {1997})}\BibitemShut {NoStop}%
\bibitem [{\citenamefont {Sadhu}\ and\ \citenamefont {Derrida}(2015)}]{Sadhua}%
  \BibitemOpen
  \bibfield  {author} {\bibinfo {author} {\bibfnamefont {T.}~\bibnamefont
  {Sadhu}}\ and\ \bibinfo {author} {\bibfnamefont {B.}~\bibnamefont
  {Derrida}},\ }\href@noop {} {\bibfield  {journal} {\bibinfo  {journal} {J.
  Stat. Mech.}\ ,\ \bibinfo {pages} {P09008}} (\bibinfo {year}
  {2015})}\BibitemShut {NoStop}%
\bibitem [{\citenamefont {Cividini}\ \emph {et~al.}(2016)\citenamefont
  {Cividini}, \citenamefont {Kundu}, \citenamefont {Majumdar},\ and\
  \citenamefont {Mukamel}}]{Cividini2016c}%
  \BibitemOpen
  \bibfield  {author} {\bibinfo {author} {\bibfnamefont {J.}~\bibnamefont
  {Cividini}}, \bibinfo {author} {\bibfnamefont {A.}~\bibnamefont {Kundu}},
  \bibinfo {author} {\bibfnamefont {S.~N.}\ \bibnamefont {Majumdar}}, \ and\
  \bibinfo {author} {\bibfnamefont {D.}~\bibnamefont {Mukamel}},\ }\href
  {\doibase 10.1088/1742-5468/2016/05/053212} {\bibfield  {journal} {\bibinfo
  {journal} {Journal of Statistical Mechanics: Theory and Experiment}\ }\textbf
  {\bibinfo {volume} {2016}} (\bibinfo {year} {2016}),\
  10.1088/1742-5468/2016/05/053212},\ \Eprint {http://arxiv.org/abs/1602.06689}
  {arXiv:1602.06689} \BibitemShut {NoStop}%
\bibitem [{\citenamefont {Ooshida}\ and\ \citenamefont
  {Otsuki}(2018)}]{Ooshida2018}%
  \BibitemOpen
  \bibfield  {author} {\bibinfo {author} {\bibfnamefont {T.}~\bibnamefont
  {Ooshida}}\ and\ \bibinfo {author} {\bibfnamefont {M.}~\bibnamefont
  {Otsuki}},\ }\href {\doibase 10.1088/1361-648X/aad4cc} {\bibfield  {journal}
  {\bibinfo  {journal} {Journal of Physics: Condensed Matter}\ }\textbf
  {\bibinfo {volume} {30}},\ \bibinfo {pages} {374001} (\bibinfo {year}
  {2018})}\BibitemShut {NoStop}%
\bibitem [{Note2()}]{Note2}%
  \BibitemOpen
  \bibinfo {note} {Note, however, that the very specific case of $\rho =1/2$
  has been derived using a mapping to the current in the SEP \cite
  {Derrida2009a, Derrida2009, Sadhua}.}\BibitemShut {Stop}%
\bibitem [{SM()}]{SM}%
  \BibitemOpen
  \href@noop {} {\bibinfo  {journal} {Supplementary Material}\ }\BibitemShut
  {NoStop}%
\bibitem [{\citenamefont {Brummelhuis}\ and\ \citenamefont
  {Hilhorst}(1989)}]{Brummelhuis1989a}%
  \BibitemOpen
\bibfield  {journal} {  }\bibfield  {author} {\bibinfo {author} {\bibfnamefont
  {M.~J. A.~M.}\ \bibnamefont {Brummelhuis}}\ and\ \bibinfo {author}
  {\bibfnamefont {H.~J.}\ \bibnamefont {Hilhorst}},\ }\href {\doibase
  10.1016/0378-4371(89)90082-4} {\bibfield  {journal} {\bibinfo  {journal}
  {Physica A}\ }\textbf {\bibinfo {volume} {156}},\ \bibinfo {pages} {575}
  (\bibinfo {year} {1989})}\BibitemShut {NoStop}%
\bibitem [{\citenamefont {Brummelhuis}\ and\ \citenamefont
  {Hilhorst}(1988)}]{Brummelhuis1988}%
  \BibitemOpen
  \bibfield  {author} {\bibinfo {author} {\bibfnamefont {M.~J. A.~M.}\
  \bibnamefont {Brummelhuis}}\ and\ \bibinfo {author} {\bibfnamefont {H.~J.}\
  \bibnamefont {Hilhorst}},\ }\href {\doibase 10.1007/BF01011556} {\bibfield
  {journal} {\bibinfo  {journal} {J. Stat. Phys.}\ }\textbf {\bibinfo {volume}
  {53}},\ \bibinfo {pages} {249} (\bibinfo {year} {1988})}\BibitemShut
  {NoStop}%
\bibitem [{\citenamefont {Oshanin}\ \emph {et~al.}(2004)\citenamefont
  {Oshanin}, \citenamefont {B{\'e}nichou}, \citenamefont {Burlatsky},\ and\
  \citenamefont {Moreau}}]{Gleb}%
  \BibitemOpen
  \bibfield  {author} {\bibinfo {author} {\bibfnamefont {G.}~\bibnamefont
  {Oshanin}}, \bibinfo {author} {\bibfnamefont {O.}~\bibnamefont
  {B{\'e}nichou}}, \bibinfo {author} {\bibfnamefont {S.~F.}\ \bibnamefont
  {Burlatsky}}, \ and\ \bibinfo {author} {\bibfnamefont {M.}~\bibnamefont
  {Moreau}},\ }in\ \href@noop {} {\emph {\bibinfo {booktitle} {Instabilities
  and Nonequilibrium Structures IX}}},\ \bibinfo {editor} {edited by\ \bibinfo
  {editor} {\bibfnamefont {O.}~\bibnamefont {Descalzi}}, \bibinfo {editor}
  {\bibfnamefont {J.}~\bibnamefont {Mart{\'i}nez}}, \ and\ \bibinfo {editor}
  {\bibfnamefont {S.}~\bibnamefont {Rica}}}\ (\bibinfo  {publisher} {Springer
  Netherlands},\ \bibinfo {address} {Dordrecht},\ \bibinfo {year} {2004})\ pp.\
  \bibinfo {pages} {33--74}\BibitemShut {NoStop}%
\bibitem [{\citenamefont {Derrida}\ and\ \citenamefont
  {Gerschenfeld}(2009{\natexlab{a}})}]{Derrida2009a}%
  \BibitemOpen
  \bibfield  {author} {\bibinfo {author} {\bibfnamefont {B.}~\bibnamefont
  {Derrida}}\ and\ \bibinfo {author} {\bibfnamefont {A.}~\bibnamefont
  {Gerschenfeld}},\ }\href {\doibase 10.1007/s10955-009-9830-1} {\bibfield
  {journal} {\bibinfo  {journal} {Journal of Statistical Physics}\ }\textbf
  {\bibinfo {volume} {137}},\ \bibinfo {pages} {978} (\bibinfo {year}
  {2009}{\natexlab{a}})},\ \Eprint {http://arxiv.org/abs/0907.3294}
  {arXiv:0907.3294} \BibitemShut {NoStop}%
\bibitem [{\citenamefont {Derrida}\ and\ \citenamefont
  {Gerschenfeld}(2009{\natexlab{b}})}]{Derrida2009}%
  \BibitemOpen
  \bibfield  {author} {\bibinfo {author} {\bibfnamefont {B.}~\bibnamefont
  {Derrida}}\ and\ \bibinfo {author} {\bibfnamefont {A.}~\bibnamefont
  {Gerschenfeld}},\ }\href {\doibase 10.1007/s10955-009-9772-7} {\bibfield
  {journal} {\bibinfo  {journal} {Journal of Statistical Physics}\ }\textbf
  {\bibinfo {volume} {136}},\ \bibinfo {pages} {1} (\bibinfo {year}
  {2009}{\natexlab{b}})},\ \Eprint {http://arxiv.org/abs/0902.2364}
  {arXiv:0902.2364} \BibitemShut {NoStop}%
\end{thebibliography}

\begin{thebibliography}{3}%
\makeatletter
\providecommand \@ifxundefined [1]{%
 \@ifx{#1\undefined}
}%
\providecommand \@ifnum [1]{%
 \ifnum #1\expandafter \@firstoftwo
 \else \expandafter \@secondoftwo
 \fi
}%
\providecommand \@ifx [1]{%
 \ifx #1\expandafter \@firstoftwo
 \else \expandafter \@secondoftwo
 \fi
}%
\providecommand \natexlab [1]{#1}%
\providecommand \enquote  [1]{``#1''}%
\providecommand \bibnamefont  [1]{#1}%
\providecommand \bibfnamefont [1]{#1}%
\providecommand \citenamefont [1]{#1}%
\providecommand \href@noop [0]{\@secondoftwo}%
\providecommand \href [0]{\begingroup \@sanitize@url \@href}%
\providecommand \@href[1]{\@@startlink{#1}\@@href}%
\providecommand \@@href[1]{\endgroup#1\@@endlink}%
\providecommand \@sanitize@url [0]{\catcode `\\12\catcode `\$12\catcode
  `\&12\catcode `\#12\catcode `\^12\catcode `\_12\catcode `\%12\relax}%
\providecommand \@@startlink[1]{}%
\providecommand \@@endlink[0]{}%
\providecommand \url  [0]{\begingroup\@sanitize@url \@url }%
\providecommand \@url [1]{\endgroup\@href {#1}{\urlprefix }}%
\providecommand \urlprefix  [0]{URL }%
\providecommand \Eprint [0]{\href }%
\providecommand \doibase [0]{http://dx.doi.org/}%
\providecommand \selectlanguage [0]{\@gobble}%
\providecommand \bibinfo  [0]{\@secondoftwo}%
\providecommand \bibfield  [0]{\@secondoftwo}%
\providecommand \translation [1]{[#1]}%
\providecommand \BibitemOpen [0]{}%
\providecommand \bibitemStop [0]{}%
\providecommand \bibitemNoStop [0]{.\EOS\space}%
\providecommand \EOS [0]{\spacefactor3000\relax}%
\providecommand \BibitemShut  [1]{\csname bibitem#1\endcsname}%
\let\auto@bib@innerbib\@empty
\bibitem [{\citenamefont {Prudnikov}\ \emph {et~al.}(1986)\citenamefont
  {Prudnikov}, \citenamefont {Brychkov},\ and\ \citenamefont
  {Marichev}}]{supp:Prudnikov_vol2}%
  \BibitemOpen
  \bibfield  {author} {\bibinfo {author} {\bibfnamefont {A.~P.}\ \bibnamefont
  {Prudnikov}}, \bibinfo {author} {\bibfnamefont {A.}~\bibnamefont {Brychkov}},
  \ and\ \bibinfo {author} {\bibfnamefont {O.~I.}\ \bibnamefont {Marichev}},\
  }\href@noop {} {\emph {\bibinfo {title} {Integrals and series: special
  functions}}},\ Vol.~\bibinfo {volume} {2}\ (\bibinfo  {publisher} {CRC
  Press},\ \bibinfo {year} {1986})\BibitemShut {NoStop}%
\bibitem [{\citenamefont {Sadhu}\ and\ \citenamefont {Derrida}(2015)}]{supp:Sadhua}%
  \BibitemOpen
  \bibfield  {author} {\bibinfo {author} {\bibfnamefont {T.}~\bibnamefont
  {Sadhu}}\ and\ \bibinfo {author} {\bibfnamefont {B.}~\bibnamefont
  {Derrida}},\ }\href@noop {} {\bibfield  {journal} {\bibinfo  {journal} {J.
  Stat. Mech.}\ ,\ \bibinfo {pages} {P09008}} (\bibinfo {year}
  {2015})}\BibitemShut {NoStop}%
\bibitem [{\citenamefont {Krapivsky}\ \emph {et~al.}(2015)\citenamefont
  {Krapivsky}, \citenamefont {Mallick},\ and\ \citenamefont
  {Sadhu}}]{supp:Krapivsky2015}%
  \BibitemOpen
  \bibfield  {author} {\bibinfo {author} {\bibfnamefont {P.~L.}\ \bibnamefont
  {Krapivsky}}, \bibinfo {author} {\bibfnamefont {K.}~\bibnamefont {Mallick}},
  \ and\ \bibinfo {author} {\bibfnamefont {T.}~\bibnamefont {Sadhu}},\ }\href
  {\doibase 10.1007/s10955-015-1291-0} {\bibfield  {journal} {\bibinfo
  {journal} {Journal of Statistical Physics}\ }\textbf {\bibinfo {volume}
  {160}},\ \bibinfo {pages} {885} (\bibinfo {year} {2015})}\BibitemShut
  {NoStop}%
\end{thebibliography}

%

\clearpage

\onecolumngrid

\begin{center}

{\large\textbf{Cumulant generating functions of a tracer\\ in quenched dense symmetric exclusion processes}}

$\ $

{\large\textbf{\textit{{Supplementary Material}}}}

$\ $

Alexis Poncet,$^1$ Olivier B\'enichou,$^1$ and Pierre Illien$^2$

$\ $

$^1$\textit{Sorbonne Universit\'e, CNRS, Laboratoire de Physique Th\'eorique de la Mati\`ere Condens\'ee (LPTMC), 4 Place Jussieu, 75005 Paris, France}

$^2$\textit{Sorbonne Universit\'e, CNRS, Laboratoire de Physico-Chimie des \'Electrolytes et Nanosyst\`emes Interfaciaux (PHENIX), 4 Place Jussieu, 75005 Paris, France}

\end{center}

\setcounter{equation}{0}
\setcounter{figure}{0}

\renewcommand{\theequation}{S\arabic{equation}}
\renewcommand{\thefigure}{S\arabic{figure}}

\renewcommand*{\citenumfont}[1]{S#1}
\renewcommand*{\bibnumfmt}[1]{[S#1]}

\section{Step of density in the dense symmetric exclusion process}
We consider the symmetric exclusion process with different initial densities in front of and behind a tracer which is initially at the origin. We call $\rho_+$ the density in front and $\rho_-$ the density behind. The average density is $\rho = (\rho_+ + \rho_-)/2$.
We will consider the dense limit $\rho\to 1$. We define the step
$\sigma = (\rho_- - \rho_+)/[2(1-\rho)]$ and remark that $1-\rho_\pm = (1-\rho)(1\pm \sigma)$.

We look at a 1D lattice of finite size $N$, with $N = N_+ + N_- + 1$. $N_+$ (resp. $N_-$) is the number of sites with strictly positive (resp. strictly negative) indices.
We consider $M_+$ vacancies at initial positions $Z_j^+ > 0$ ($j=1, \dots M_+$) and $M_-$ vacancies at initial positions $Z_j^- < 0$ ($j=1, \dots M_-$). The densities are such that $M_\pm / N_\pm = 1-\rho_\pm$.
One checks that Eq.~(2) of the main text is modified into
\begin{equation} \label{smeq:lim}
\tilde P^{(t)} (k|\{Z_j^+\}, \{Z_j^-\}) \underset{\rho \to 1}{\sim} \prod_{j=1}^{M_+} \tilde p^{(t)}_{Z_j^+}(k)\prod_{j=1}^{M_-} \tilde p^{(t)}_{Z_j^-}(k).
\end{equation}

We now investigate the two cases studied in the article: annealed and quenched initial conditions.

\subsection{Annealed initial conditions}
In the case of annealed initial conditions, the average propagator (Eq.~(4) of the main text) is now defined as
\begin{align}
\tilde P_A^{(t)} (k) &\equiv \frac{1}{(N_+)^{M_+} (N_-)^{M_-}} \sum_{Z_1^+, \dots, Z_{M_+}^+>0}\sum_{Z_1^-, \dots, Z_{M_-}^-<0}\tilde P^{(t)} (k|\{Z_j^+\}, \{Z_j^-\}).
\end{align}
We use Eq.~\eqref{smeq:lim} and quickly obtain
\begin{align}
\tilde P_A^{(t)} (k) &\underset{\rho \to 1}{\sim}
\left[\frac{1}{N^+} \sum_{Z> 0} \tilde p^{(t)}_{Z}(k)\right]^{M_+}
\left[\frac{1}{N^-} \sum_{Z< 0} \tilde p^{(t)}_{Z}(k)\right]^{M_-} \\
&\underset{\rho \to 1}{\sim} \left[1 + \frac{1}{N^+} \sum_{Z> 0} \left( \tilde p^{(t)}_{Z}(k) - 1\right)\right]^{M_+}
\left[1 + \frac{1}{N^-} \sum_{Z< 0} \left( \tilde p^{(t)}_{Z}(k) - 1\right)\right]^{M_-} .
\end{align}
We now consider the large size limit $M_\pm, N_\pm\to\infty$ at constant densities: $M_+/N_+ = 1-\rho_+ = (1-\rho)(1+\sigma)$ and $M_-/N_- = 1-\rho_- = (1-\rho)(1-\sigma)$. This gives us an expression for the annealed cumulant-generating function at high density:
\begin{align}
\lim_{\rho \to 1} \frac{\psi_A^{(t)}(k)}{1-\rho} =
\lim_{\rho \to 1} \frac{\log \tilde P_A^{(t)} (k)}{1-\rho}
= \sum_{Z\neq 0} \left(\tilde p^{(t)}_{Z}(k)-1\right)
+ \sigma \sum_{Z>0} \left(\tilde p^{(t)}_{Z}(k)-\tilde p^{(t)}_{-Z}(k)\right).
\end{align}

Using the Laplace transform of $\tilde p^{(t)}_{Z}(k)$ [Eq.~\eqref{psingle_real} of the main text], we obtain
\begin{equation}
\lim_{\rho \to 1} \frac{\hat \psi_A(k, \xi)}{1-\rho}
\underset{\xi\to 1}{\sim} \frac{1}{\sqrt{2}}\frac{1}{(1-\xi)^{3/2}}
\left\{p_1 (1+\sigma)(e^{ik} - 1) + p_{-1}(1-\sigma)(e^{-ik} - 1) \right\}.
\end{equation}
This gives us the following large-time scaling:
\begin{equation}
\lim_{\rho \to 1} \frac{\psi_A(k, t)}{1-\rho}
\underset{\xi\to 1}{\sim} \sqrt\frac{2t}{\pi}
\left\{p_1 (1+\sigma)(e^{ik} - 1) + p_{-1}(1-\sigma)(e^{-ik} - 1) \right\}.
\end{equation}
In the case of symmetric jumps, $p_1 = p_{-1} = 1/2$, we obtain the cumulants given in Eqs.~\eqref{K_odd_step_ann} and \eqref{K_even_step_ann} of the main text.

\subsection{Quenched initial conditions}
In the case of quenched initial conditions, the average of the cumulant-generating function (Eq.~\eqref{psiQaverage} of the main text) is now defined as
\begin{equation}
\psi_Q(k, t) \equiv \frac{1}{N_+^{M_+} N_-^{M_-}} \sum_{Z^+_1, \dots, Z^+_{M_+} > 0} \sum_{Z_1^-, \dots, Z^-_{M_-} < 0}
\psi(k, t|\{Z_j^+\}, \{Z_j^-\})
\end{equation}
with $\psi(k, t|\{Z_j^+\}, \{Z_j^+\}) \equiv \log\tilde P^{(t)} (k|\{Z_j^+\}, \{Z_j^-\}) $. We use Eq.~\eqref{smeq:lim} and obtain

\begin{equation}
\lim_{\rho \to 1} \frac{\psi_Q(k, t)}{1-\rho} = (1+\sigma) \sum_{Z>0} \log \tilde p^{(t)}_{Z}(k) + (1-\sigma) \sum_{Z< 0} \log \tilde p^{(t)}_{Z}(k).
\end{equation}
We now consider the expression of $\tilde p^{(t)}_{Z}(k)$ given in Eq.~\eqref{psingle_real} of the main text. Performing the Riemann summation, we obtain the large time scaling of the quenched cumulant-generating function:
\begin{equation}
\lim_{\rho \to 1} \frac{\psi_Q(k, t)}{1-\rho} \underset{t\to\infty}{\sim}
\sqrt{2t} \left[(1+\sigma)\int_0^\infty \dd z \,   \log \left[1 + p_1 \left(\ex{\ii k} -1\right)
\text{erfc} z\right]
+(1-\sigma) \int_0^\infty \dd z \,   \log
\left[1 + p_{-1} \left(\ex{-\ii k} -1\right)
\text{erfc} z\right]
\right].
\end{equation}
This is the result that we report in Eq.~\eqref{eq:1tp_resPsiQ_step} of the main text, in the case of symmetric jumps ($p_1 = p_{-1} = 1/2$).

\section{Definite integrals of powers of the complementary error function}
We recall some results from Ref.~\cite{supp:Prudnikov_vol2} (2.8.20) that are useful to compute the quenched cumulants up to order 4.
\begin{align}
\int_0^\infty \dd z \erfc z &= \frac{1}{\sqrt\pi} \\
\int_0^\infty \dd z \erfc^2 z &= \frac{2-\sqrt{2}}{\sqrt\pi}\\
\int_0^\infty \dd z \erfc^3 z &= \frac{3}{\pi^{3/2}} \left[
\pi(1-\sqrt{2}) + 2\sqrt{2} \arctan\frac{\sqrt{2}}{2}\right] \\
\int_0^\infty \dd z \erfc^4 z &= \frac{2}{\pi^{3/2}} \left[
\pi(2-3\sqrt{2}) + 6\sqrt{2}\left(2\arctan\frac{\sqrt{2}}{2} - \arctan\frac{\sqrt{2}}{4}\right)\right]
\end{align}

\section{Numerical simulations}
\subsection{Uniform density}
The simulations requiring an uniform density (Fig. 2) are performed with periodic boundary conditions. The size of the system is $N=1000$ and the density is $0.95$. Therefore, there are $K=950$ particles and $50$ vacancies. Initially, the vacancies are on the sites $20(k+1/2)$ for $0 \leq k \leq 49$ (a density of $0.98$ is achieved in a similar way). We monitor the position of the particle initially at the origin (tracer).

The time evolution of the system is achieved in the following way. We draw a random number according to an exponential law of rate $1/K$, it corresponds to the time increment. We then choose a particle uniformly at random and attempt to move it to its left or right neighboring site with equal probabilities (for the bath particle) or according its jump probabilities  (for the tracer). If the arrival site is empty, the jump is done, otherwise it is rejected.

The moments of the tracer are recorded every time interval $\Delta t = 1$ and they are averaged over $2\cdot 10^6$ simulations. In Figs. 2 and 3 the data is averaged over bins of equal size in logarithmic time.

\subsection{Step of density}
The simulations requiring an step of density are performed using open boundary conditions: there is a reservoir at density $\rho_-$ on the left and a reservoir at density $\rho_+$ on the right. The size of the system is $N=10000$, it is large enough for the influence of the reservoirs (which are annealed, by definition) to be negligible.
Initially, all the sites are occupied except those at positions $\ell_+ (k+1/2)$ and $-\ell_-(k+1/2)$ for $k\geq 0$, with $\ell_+ = (1-\rho_+)^{-1}$ and $\ell_- = (1-\rho_-)^{-1}$.

The time evolution of the system is implemented by a Gillespie algorithm similar to the one described for uniform density. The additional ingredients are (i) the left reservoir is associated with a rate $\rho_-/2$ at which a particle is created on the leftmost site if it is empty; (ii) if a particle on the leftmost site tries to jump to the left, it destroyed with probability $(1-\rho_-)$; and \textit{mutatis mutandis} for the right reservoir.

The moments of the tracer are recorded every time interval $\Delta t = 1$ and they are averaged over $2\cdot 10^4$ simulations. In Fig. 3 the data is averaged over bins of equal size in logarithmic time.

\section{Approximate expression in the symmetric case}
\begin{figure*}
	\centering
	\includegraphics[scale=1]{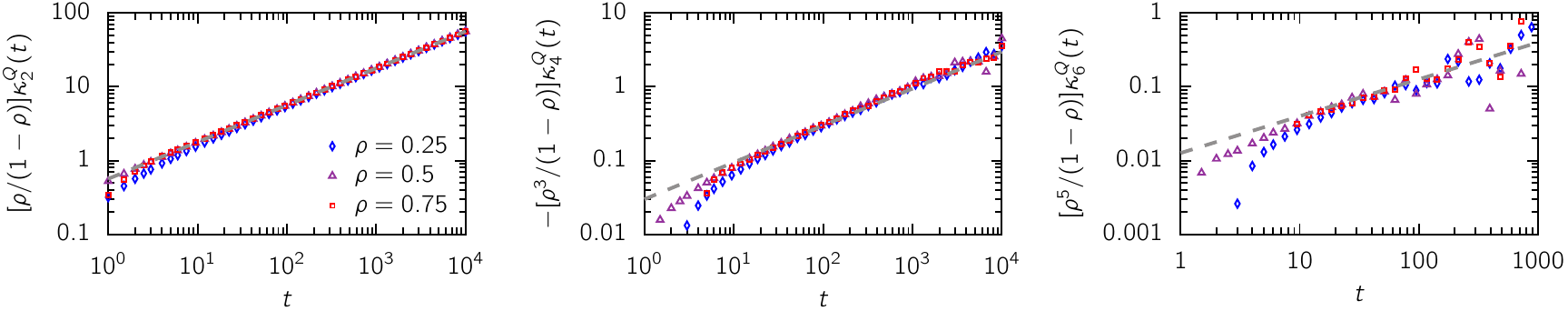}
	\caption{Quenched cumulants $\kappa_2^Q$, $\kappa_4^Q$ and $\kappa_6^Q$ from numerical simulations at densities $0.25$, $0.5$ and $0.75$ (blue to red). The dashed gray line corresponds to the approximate predictions from Eqs.~\eqref{K2_Q_arbitrary_dens}-\eqref{K6_Q_arbitrary_dens}.}
	\label{fig:approx}
\end{figure*}

\begin{figure*}
	\centering
	\includegraphics[scale=1]{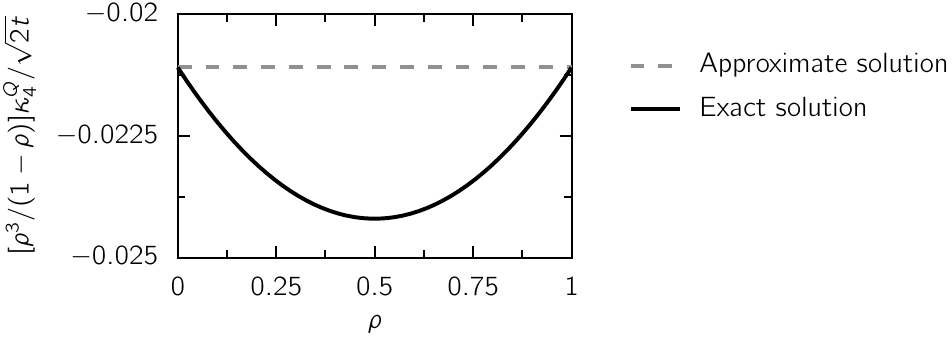}
	\caption{Comparison between the approximate solution (Eq.~\eqref{K4_Q_arbitrary_dens}, dashed gray line) and the exact solution (Eq.~\eqref{K4_Q_exact}, continuous black line) for the prefactor of the fourth quenched cumulant $\kappa_4^Q$.
	We plot $[\rho^3/(1-\rho)]\kappa_4^Q / \sqrt{2t}$ and observe that the maximum deviation of the approximate prefactor remains small.}
	\label{fig:compa}
\end{figure*}

At the end of the article, we put forward the following approximate expression for the CGF of a tagged particle in the symmetric SEP with quenched initial conditions,
\begin{align}
&\psi^\text{approx}_Q(k, t)\underset{t\to\infty}{\sim}\rho(1-\rho)\sqrt{2t}
\int_0^\infty dz\log\left[1 - \sin^2\left(\frac{k}{2\rho}\right) \text{erfc}( z) \text{erfc}(-z)\right].
\label{supp:CGF_sym_guess}
\end{align}
This expression is exact in the limits $\rho\to 0$~\cite{supp:Sadhua} and $\rho\to 1$ (this article).
It leads to the following expressions for the lowest cumulants,
\begin{align}
\kappa_2^Q(t)\underset{t\to\infty}{\sim}& \frac{1}{\sqrt{2\pi}} \frac{1-\rho}{\rho} \sqrt{2t}, \label{K2_Q_arbitrary_dens} \\
\kappa_4^{Q, \text{approx}}(t)\underset{t\to\infty}{\sim}& \sqrt\frac{2}{\pi}\left(\frac{9}{\pi} \arctan\frac{1}{2\sqrt{2}} - 1\right)\frac{1-\rho}{\rho^3} \sqrt{2t},  \label{K4_Q_arbitrary_dens} \\
\kappa_6^{Q, \text{approx}}(t)\underset{t\to\infty}{\sim}& 
\left[\int_0^\infty \dd z  \frac{15}{4} (1-\erf^2 z)\left(\left(\erf^2 z-\frac{1}{2}\right)^{2}-\frac{7}{60}\right)\right] \frac{1-\rho}{\rho^5} \sqrt{2t}. \label{K6_Q_arbitrary_dens}
\end{align}
These expressions look in good agreement with numerical simulations
(Fig.~\ref{fig:approx}) which could points towards the exact nature of this expression. Unfortunately, the fourth cumulant has been computed rigorously~\cite{supp:Krapivsky2015} and
reads
\begin{equation}
\kappa_4^{Q, \text{exact}} \underset{t\to\infty}{\sim} \sqrt\frac{2}{\pi}
\left\{(1-2\rho)^2 \left[\frac{9}{\pi} \arctan\left(\frac{1}{2\sqrt{2}}\right) - 1\right]
+ \rho(1-\rho) \left(2-\frac{3}{\sqrt{2}}\right)
\right\}
\frac{1-\rho}{\rho^3} \sqrt{2t} \label{K4_Q_exact}
\end{equation}
As expected, the exact and approximate expressions are equal in both limits $\rho\to 0$ and $\rho\to 1$. Furthermore we show on Fig.~\ref{fig:compa} that the difference between the prefectors from the exact and approximate expressions is always less than 15~\% so that the approximate expression that we put forward is reasonnable.


\bibliographystyle{apsrev4-1}

\end{document}